\documentclass[numreferences]{kapproc} 
 \normallatexbib

\usepackage{amssymb}
\usepackage[dvips]{graphicx}
\begin{document}

\articletitle{Bipolaronic proximity  and other unconventional
effects in cuprate superconductors}

\author{A. S. Alexandrov}
\affil{Department of Physics, Loughborough University,\\
Loughborough, United Kingdom} \email{a.s.alexandrov@lboro.ac.uk}

\begin{abstract}
\small {There is   compelling evidence for a strong electron-phonon
interaction (EPI) in cuprate superconductors  from the isotope
effects on the supercarrier mass, high resolution angle resolved
photoemission spectroscopies (ARPES),  a number of optical and
neutron-scattering measurements in accordance with our prediction of
high-temperature superconductivity in polaronic liquids. A number of
observations point to the possibility that high-$T_{c}$ cuprate
superconductors may not be conventional Bardeen-Cooper-Schrieffer
(BCS) superconductors, but rather derive from the Bose-Einstein
condensation (BEC) of real-space pairs, which are mobile small
bipolarons. Here I  review the bipolaron theory  of unconventional
 proximity effects, the symmetry and
checkerboard modulations of the order parameter and  quantum
magneto-oscillations discovered recently in cuprates.}
\end{abstract}

\begin{keywords}
 bipolarons, cuprates, proximity, symmetry, magnetooscillations
\end{keywords}

\section{Polarons in high-temperature superconductors}
\label{evidence}

Many unconventional  properties of cuprate superconductors  may be
attributed to the Bose-Einstein condensation (BEC) of real-space
pairs, which are mobile small bipolarons
\cite{alexandrov:1988,alemot,dev,ale5,alebook,edw,alenar,alejpcm}. A
possible fundamental origin of such strong departure of the cuprates
from conventional BCS behaviour is  the unscreened Fr\"ohlich EPI
providing the polaron level shift $E_p$ of the order of 1 eV
\cite{ale5}, which is routinely neglected in the Hubbard $U$ and
$t-J$ models. This huge interaction with $c-$axis polarized optical
phonons is virtually unscreened  at any doping of cuprates. Even
acting alone it leads to large bipolarons in the continuum limit, if
the ratio $\eta = \epsilon/\epsilon_0$ of the high-frequency
(electronic) and static dielectric constants is small enough
\cite{dev}. The large  Fr\"ohlich bipolarons are further stabilised
in going from 3D to 2D \cite{verbist}. When acting together with the
deformation potential and the Jahn-Teller EPIs the  Fr\"ohlich EPI
overcomes the inter-site Coulomb repulsion forming small bipolarons
\cite{alebook}. Hence, in order to build the adequate theory of
high-temperature superconductivity, the  Coulomb repulsion and the
\emph{unscreened} long-range EPI together with the short-range one
should be treated on an equal footing with the short-range repulsive
Hubbard $U$. When these interactions are strong compared with the
kinetic energy of carriers, this "Coulomb-Fr\"ohlich" model predicts
the ground state in the form of superlight small bipolarons
\cite{ale5,alekor,hague}.

Nowadays  compelling evidence for a strong EPI  has arrived from
isotope effects  \cite{zhao,ref2}, more recent high resolution angle
resolved photoemission spectroscopies (ARPES)
 \cite{LAN}, and a number of earlier optical
 \cite{mic1,ita,tal,tim,devopt},
 neutron-scattering \cite{ega} and recent  inelastic scattering studies \cite{rez}
 of cuprates and related compounds.
 Whereas calculations based on the local spin-density
approximation (LSDA) often predict negligible EPI, the inclusion of
Hubbard $U$ in the $LSDA+U$ calculations greatly enhances its
strength \cite{zha}.

A parameter-free estimate of the Fermi energy using the
magnetic-field penetration depth found a very low value, $\epsilon
_{F}\lesssim 100$ meV \cite{alexandrov:2001b} clearly supporting the
real-space (i.e individual) pairing in cuprate superconductors.
There is strong experimental evidence for a gap in the normal-state
electron density of states of cuprates \cite{alemot}, which is known
as the pseudogap. Experimentally measured pseudogaps of many
cuprates are  $ \gtrsim 50$ meV \cite{kabmic}. If following Ref.
\cite{alexandrov:1991} one accepts
 that the pseudogap is about half of the pair
binding energy, $\Delta$,  then the condition for real-space
pairing, $\epsilon _{F}\lesssim \pi \Delta$, is well satisfied in
most cuprates (typically the small bipolaron radius is $r_b\approx
0.2 -0.4$ nm).

Also  magnetotransport  and thermal magnetotransport
 data strongly support preformed bosons in cuprates.
In particular,
 many high-magnetic-field studies revealed a non-BCS upward
curvature of the upper critical field $H_{c2}(T)$ (see \cite{ZAV}
for a review of experimental data),  in accordance with the
theoretical prediction for the Bose-Einstein condensation of charged
bosons in the magnetic field \cite{aleH}. The Lorenz
number, $L= e^{2}\kappa _{e}/T\sigma $ differs significantly from the Sommerfeld value $%
L_{e}=\pi ^{2}/3$ of the standard Fermi-liquid theory, if carriers
are double-charged bosons \cite{NEV}. Here $\kappa _{e}$, and
$\sigma $  are  electron thermal and electrical conductivities,
respectively. Ref. \cite{NEV} predicted a rather low Lorenz number
for bipolarons, $L\approx 0.15L_{e}$, due to the double elementary
charge of bipolarons, and also due to their nearly classical
distribution function above $T_{c}$. Direct measurements of the
Lorenz number  using the thermal Hall effect \cite{ZHANG} produced
the value of $L$ just above $T_{c}$ about the same as predicted by
the bipolaron model,  and its strong temperature dependence. This
breakdown of the Wiedemann-Franz law is apparently caused by excited
single polarons coexisting with bipolarons in the thermal
equilibrium \cite{leeale,alelor}. Also unusual normal state
diamagnetism uncovered by  torque magnetometery
 has been convincingly explained as the normal state
(Landau) diamagnetism of
 charged bosons \cite{aledia}.

However, despite clear evidence for the existence of polarons in
cuprates, no consensus currently exists  concerning the microscopic
mechanism of high-temperature superconductivity. While a number of
early (1990s) and more recent studies prove that the Mott-Hubbard
insulator promotes doping-induced polaron formation \cite{polarons},
some other works  suggest that EPI  does not only not help, but
hinder the pairing instability. The controversy should be resolved
experimentally. Here I argue that the giant (GPE) and nil (NPE)
proximity effects  provide another piece of evidence for bipolaronic
BEC in cuprates  \cite{proxi}. The same bipolaronic scenario also
explains the symmetry and checkerboard modulations of the order
parameter, and recently observed magnetooscillations in the vortex
state of cuprates.

\section{Unconventional proximity effects}

 Several groups \cite{bozp} reported
that in the Josephson cuprate $SNS$ junctions supercurrent can run
through normal $N$-barriers with the thickness $2L$ greatly
exceeding the coherence length, when the barrier is made from a
slightly doped non-superconducting cuprate (the so-called $N'$
barrier), Fig.1.
\begin{figure}
\begin{center}
\includegraphics[angle=-90,width=0.90\textwidth]{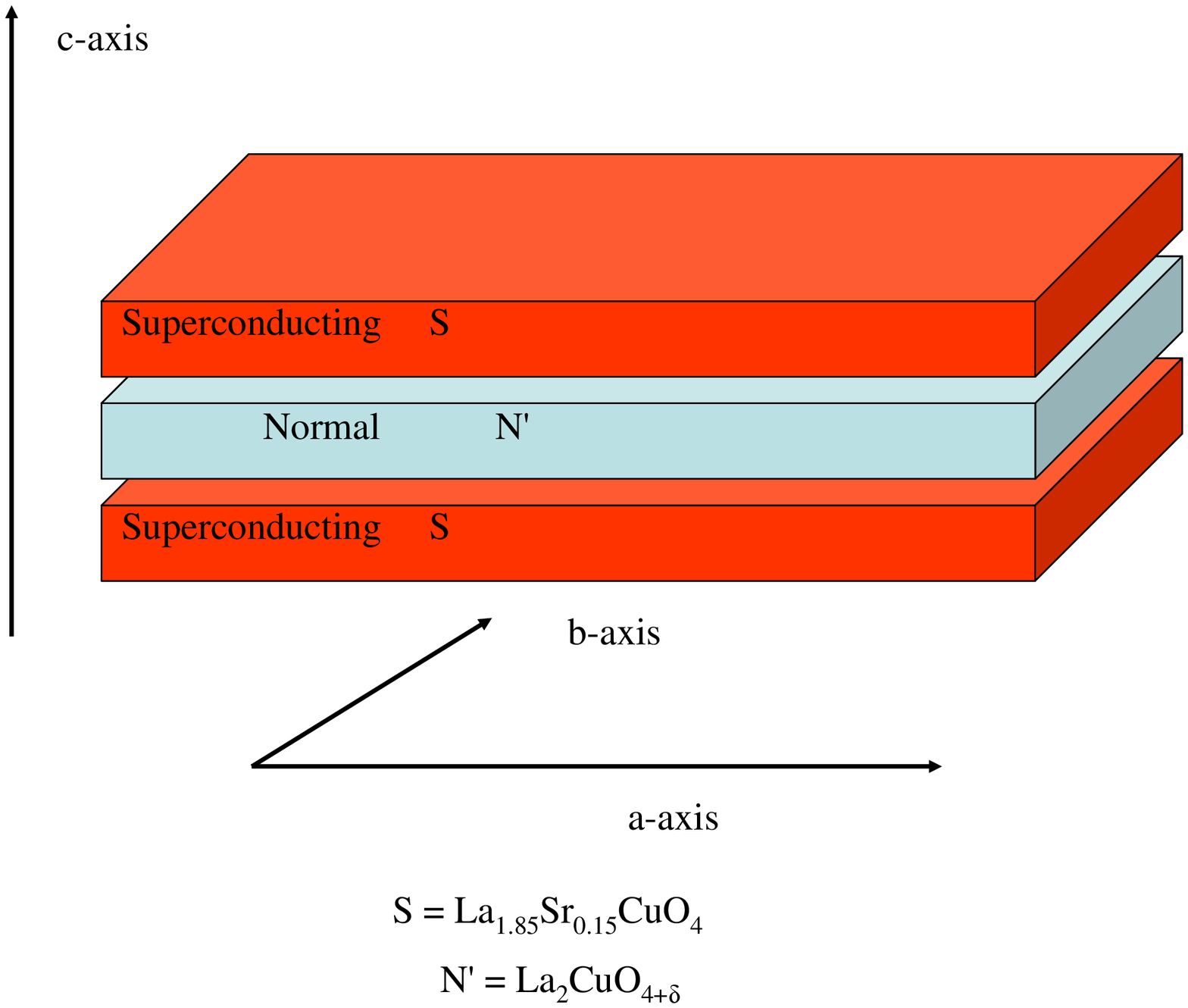}
\vskip -0.5mm \caption{SNS cuprate nanostructure}
\end{center}
\end{figure}

Using the advanced molecular beam epitaxy, Bozovic \emph{et al.}
\cite{bozp} proved that GPE is intrinsic, rather than caused by any
extrinsic inhomogeneity of the barrier. Resonant scattering of
soft-x-ray radiation did not find any signs of intrinsic
inhomogeneity (such as charge stripes, charge-density waves, etc.)
either \cite{boz5}. Hence GPE defies the conventional explanation,
which predicts that the critical current should exponentially decay
with the characteristic length of about the coherence length, $\xi
\lesssim 1$ nm in  cuprates. Annealing the junctions at low
temperatures in vacuum rendered the barrier insulating. Remarkably
when  the $SN'S$ junction was converted into a
superconductor-insulator-superconductor (SIS) device no supercurrent
was observed, even in devices with the thinnest (one unit cell
thick) barriers \cite{boz0} (nil proximity effect, NPE).

Both GPE and NPE can be broadly understood as the bipolaronic
Bose-condensate tunnelling into a cuprate \emph{semiconductor}
\cite{proxi}.

To illustrate the point one can apply the Gross-Pitaevskii (GP)-type
equation for the superconducting order parameter $\psi({\bf r})$,
generalized by us  \cite{alevor} for a charged Bose liquid (CBL),
since, as discussed above,  many observations including a small
coherence length point to a possibility that cuprate superconductors
may not be conventional BCS superconductors, but rather derive from
BEC of real-space pairs, such as  mobile small bipolarons
\cite{ale5},
\begin{equation}
\left[E(-i\hbar {\bf \nabla}+2e {\bf A})-\mu+\int d{\bf r'} V({\bf
r} -{\bf r'})|\psi ({\bf r'})|^2 \right]\psi({\bf r}) =0.\label{gp}
\end{equation}
Here $E({\bf K})$ is the center-of-mass pair dispersion and the
Peierls substitution, ${\bf K} \Rightarrow -i\hbar {\bf \nabla}+2e
{\bf A}$ is applied with the vector potential ${\bf A}({\bf r})$.

The integro-differential equation (\ref{gp})
 is quite different from the  Ginzburg-Landau \cite{gin}
and  Gross-Pitaevskii \cite{gro} equations, describing the order
parameter in the BCS and neutral superfluids, respectively.   Here
$\mu$ is  the chemical potential and $V({\bf r})$ accounts for
 the long-range Coulomb and a short-range composed boson-boson repulsions
 \cite{alebook}. While the electric field potential can be found from the
corresponding Poisson-like equation \cite{alevor}, a solution of two
coupled nonlinear differential equations for the order parameter
$\psi({\bf r})$ and for the potential $V({\bf r})$ in the
nanostructure, Fig.1, is a nontrivial mathematical problem. For more
transparency we restrict our analysis in this section by a
short-range potential, $V({\bf r})=v |\psi({\bf r})|^2$, where a
constant $v$ accounts for the short-range repulsion. Then in the
absence of the magnetic field Eq.(\ref{gp}) is reduced to the
familiar GP equation \cite{gro} using  the continuum (effective
mass) approximation, $E({\bf K})= K^2/2m_c$. In the tunnelling
geometry of $SN'S$ junctions, Figs.1,2, it takes the form,
\begin{equation}
{1\over{2m_c}}{d^2\psi(Z)\over{dZ^2}}=[v |\psi(Z)|^2-\mu]\psi(Z),
\end{equation}
in the superconducting region, $Z<0$,  Fig.2. Here  $m_c$ is the
boson mass in the direction of tunnelling along $Z$ ($\hbar=c=k_B=1$
in this section). Deep inside the superconductor the order parameter
is a constant, $|\psi(Z)|^2=n_s$ and $\mu=vn_s$ , where the
condensate density $n_s$ is about  $n_s\approx x/2$, if the
temperature is well below $T_c$ of the superconducting electrode.
 Here the in-plane lattice constant $a$ and the unit cell volume are
 taken as unity, and $x$ is the doping level as in La$_{2-x}$Sr$_{x}$CuO$_4$.

\begin{figure}
\begin{center}
\includegraphics[angle=-90,width=0.90\textwidth]{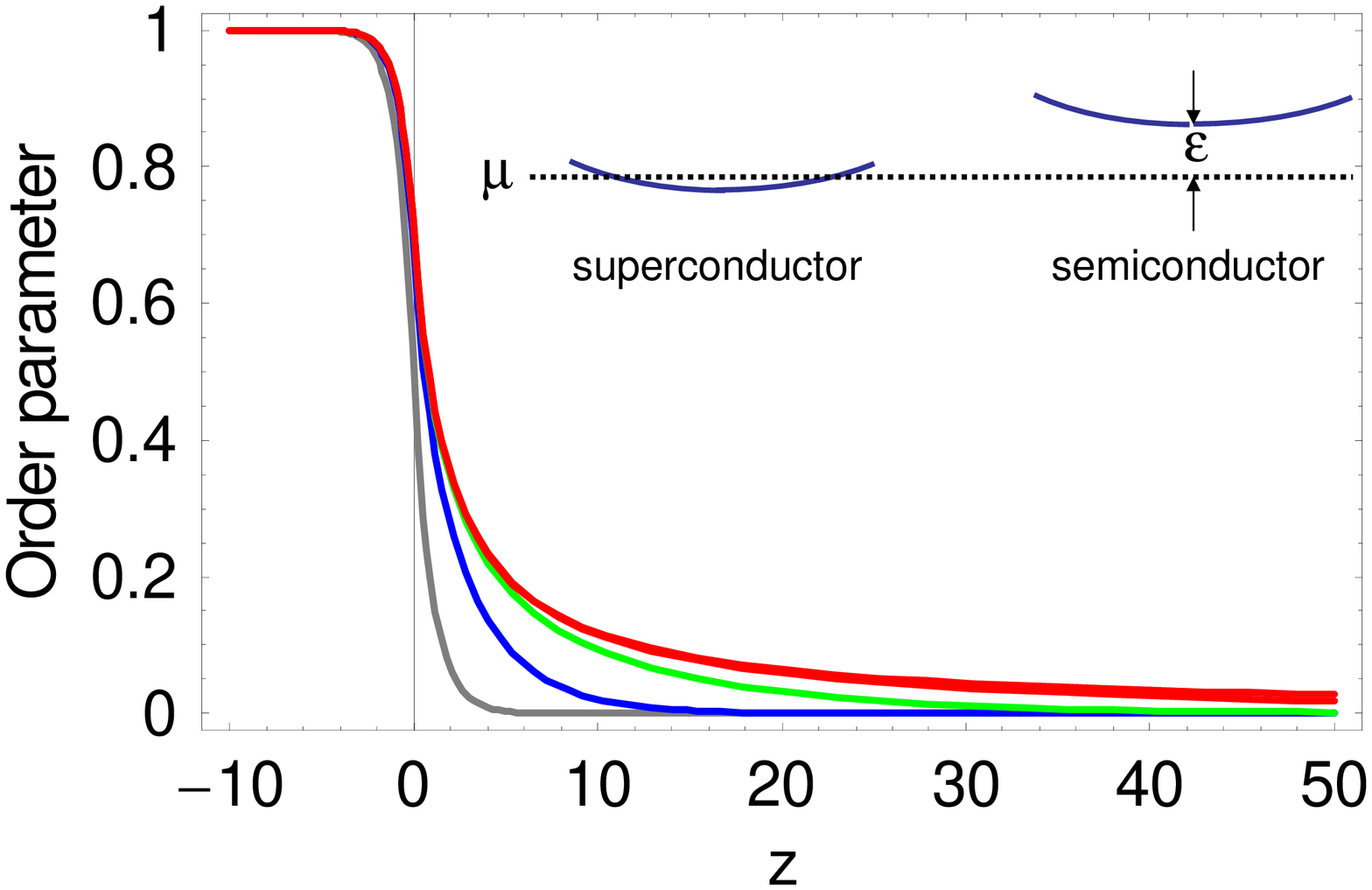}
\vskip -0.5mm \caption{BEC order parameter at the $SN$ boundary for
$\tilde{\mu}=1.0,0.1,0.01$ and $ \leqslant 0.001$ (upper curve). The
chemical potential is found above the boson band-edge due to the
boson-boson repulsion in  cuprate superconductors and below the edge
in   cuprate semiconductors with low doping.}
\end{center}
\end{figure}

The normal barrier  at $Z
>0$ is an underdoped cuprate above its
transition temperature, $T'_c<T$ where the chemical potential $\mu$
lies below the bosonic band by some energy $\epsilon$, Fig.2, found
from $ \int dE N(E)[\exp((E+\epsilon)/T)-1]^{-1}=x'/2$. Here $N(E)$
is the bipolaron density of states (DOS), and $x' < x$ is the doping
level of the barrier. In-plane bipolarons are quasi-two dimensional
repulsive bosons, propagating along the $CuO_2$ planes with the
effective mass $m$ several orders of magnitude smaller than their
out-of-plane mass, $m_c\gg m$,  \cite{alebook}. Using bipolaron band
dispersion, $E({\bf K})=K^{2}/2m+2t_{c}[1-\cos(K_{\perp}d)]$, the
density of states is found as $N(E)=(m/2\pi^2)\arccos(1-E/2t_c)$ for
$0<E<4t_{c}$, and $N(E)=m/2\pi$
 for $4t_c<E$. Here $K$ and $K_\perp$ are the in-plane and out-of-plane center-of-mass momenta, respectively, $t_c= 1/2m_cd^2$, and  $d$ is the inter-plane
 distance. As a result
one obtains
\begin{equation}
\epsilon(T)\leqslant -T\ln(1-e^{-T_0/T}),
\end{equation}
which is exponentially small at $T'_c < T \ll T_0$ turning into zero
at $T=T'_c$, where  $T'_c \approx T_0/\ln(T_0/2t_c)$, and $T_0=\pi
x'/m \gg T'_c \gg t_c$.

It is important to note that $\epsilon(T)$ remains also
 small at $T'_c (2D)\leqslant  T \ll T_0$   in the
purely two-dimensional  repulsive Bose-gas \cite{kagan}.  While in
two dimensions Bose condensation does not occur in either the ideal
or the interacting system, there is a phase transition to a
superfluid state at $T'_c(2D)= T_0/\ln (1/f_0) \ll T_0$, where
$f_0\ll 1$ depends on  the density of hard-core dilute bosons and
their repulsion \cite{popov,kagan}. The superfluid transition takes
place only if there is a residual repulsion between  bosons, i.e.
$T'_c(2D)=0$ for the ideal 2D Bose-gas. Actually $T'_c(2D)$ gives a
very good estimate for the exact Berezinski-Kosterlitz-Thouless
(BKT) critical temperature in the dilute Bose gas, where the BKT
contribution of vortices is important only very close to $T'_c(2D)$
\cite{popov}.

The GP equation in the barrier can be written as
\begin{equation}
{1\over{2m_c}}{d^2\psi(Z)\over{dZ^2}} =[v
|\psi(Z)|^2+\epsilon]\psi(Z).
\end{equation}
Introducing the bulk coherence length, $\xi= 1/(2m_c n_sv)^{1/2}$
and dimensionless $f(z)=\psi(Z)/n_s^{1/2}$,
$\tilde{\mu}=\epsilon/n_sv$, and $z=Z/\xi$ one obtains
 for a real
$f(z)$
\begin{equation}
{d^2f\over{dz^2}} =f^3-f,
\end{equation}
if $z<0$, and
\begin{equation}
{d^2f\over{dz^2}}=f^3+\tilde{\mu}f,
\end{equation}
if
 $z>0$. These equations can be readily solved using  first
integrals of motion respecting the boundary conditions,
$f(-\infty)=1$, and $f(\infty)=0$,
\begin{equation}
{df\over{dz}}= -(1/2+f^4/2-f^2)^{1/2},
\end{equation}
and
\begin{equation}
{df\over{dz}}= -(\tilde{\mu}f^2+f^4/2)^{1/2},
\end{equation}
for $z<0$ and $z>0$, respectively. The solution in the
superconducting electrode is given by
\begin{equation}
f(z)=\tanh \left[-2^{-1/2}z+0.5
\ln{{2^{1/2}(1+\tilde{\mu})^{1/2}+1}\over{2^{1/2}(1+\tilde{\mu})^{1/2}-1}}\right].
\end{equation}
It decays  in the close vicinity of the barrier from 1 to
$f(0)=[2(1+\tilde{\mu})]^{-1/2}$ in the interval about the coherence
length $\xi$. On the other side  of the boundary, $z>0$, it is given
by
\begin{equation}
f(z)={(2\tilde{\mu})^{1/2}\over{\sinh\{z\tilde{\mu}^{1/2}+\ln[2(\tilde{\mu}(1+\tilde{\mu}))^{1/2}+(1+4\tilde{\mu}(1+\tilde{\mu}))^{1/2}]\}}}
.
\end{equation}
Its profile is shown in Fig.2. Remarkably, the order parameter
penetrates  the normal layer up to the length $Z^* \thickapprox
(\tilde{\mu})^{-1/2}\xi$, which could be larger than $\xi$ by many
orders of magnitude,  if $\tilde{\mu}$ is  small. It is indeed the
case, if the barrier layer is sufficiently doped. For example,
taking $x'=0.1$,   c-axis $m_c=2000 m_e$, in-plane $m=10 m_e$,
$a=0.4$ nm, and $\xi=0.6$ nm, yields
 $T_0\approx 140$ K and $(\tilde{\mu})^{-1/2}\gtrsim 50$ at $T=25$K. Hence the
 order parameter could penetrate  the normal cuprate semiconductor
 up to  a hundred coherence lengths or even more as observed (GPE) \cite{bozp}. If the thickness of the barrier $L$ is small compared with $Z^*$,
and $(\tilde{\mu})^{1/2}\ll 1$, the order parameter decays following
 the power law, rather than exponentially,
\begin{equation}
f(z)={\sqrt{2}\over{z+2}}.
\end{equation}
Hence, for $L \lesssim Z^*$, the critical current should also decay
following the power law rayher than exponentially. On the other
hand, for the \emph{undoped}
 barrier $\tilde{\mu}$ becomes
 larger than unity, $\tilde{\mu}\varpropto \ln(mT/\pi x')\rightarrow \infty$ for any finite temperature $T$  when $x' \rightarrow
 0$, and the current should exponentially decay with the characteristic length  smaller that $\xi$, which is experimentally observed as well (NPE) \cite{boz0}.

\section{Quantum
magneto-oscillations, d-wave symmetry and  checkerboard modulations}

Until recently  no convincing signatures of
 quantum magneto-oscillations have been found  in
the normal state of cuprate superconductors despite significant
experimental efforts. There are no normal state oscillations even in
high quality single crystals of overdoped cuprates like
Tl$_2$Ba$_2$CuO$_6$, where conditions for
 de Haas-van Alphen (dHvA)  and Shubnikov-de Haas (SdH)
oscillations seem to be perfectly satisfied \cite{mac} and a large
Fermi surface is identified in the angle-resolved photoemission
spectra (ARPES)  \cite{plate}. The recent observations of
magneto-oscillations in  kinetic \cite{ley,ban} and magnetic
\cite{sin,proust} response functions of underdoped
YBa$_2$Cu$_3$O$_{6.5}$ and YBa$_2$Cu$_4$O$_8$ are perhaps even more
striking since many probes of underdoped cuprates including ARPES
\cite{shen} clearly point to a non Fermi-liquid normal state. Their
description  in the framework of the standard theory for a metal
\cite{schoen} has led to  a very small  Fermi-surface area of  a few
percent of the first Brillouin zone \cite{ley,ban,sin,proust}, and
to a
 low Fermi energy of only about \emph{the room temperature} \cite{sin}.
Clearly such oscillations are incompatible with the first-principle
(LDA) band structures of cuprates, but might be compatible with a a
low Fermi energy and non-adiabatic polaronic normal state of
charge-transfer Mott insulators as discussed in section 1.

Nevertheless one can raise a doubt concerning their normal state
origin. The magnetic length, $\lambda \equiv (\pi
\hbar/eB)^{1/2}\gtrsim 5$ nm, remains larger than the
zero-temperature in-plane coherence length, $\xi \lesssim 2$ nm,
measured independently, in  any field reached in Ref.
\cite{ley,ban,sin,proust}. Hence the magneto-oscillations are
observed in the vortex (mixed) state well below the upper critical
field, rather than in the normal state, as also confirmed by the
\emph{negative} sign of the Hall resistance \cite{ley}. It is well
known, that in "YBCO" the Hall conductivities of vortexes and
quasiparticles have opposite sign causing the sign change in the
Hall effect in the mixed state \cite{harris}. Also there is a
substantial magnetoresistance \cite{ban}, which is a signature of
the flux flow regime rather than of the normal state. Hence it would
be rather implausible if such oscillations  have a normal-state
origin due to
 small electron Fermi surface pockets \cite{proust}  with the
characteristic wave-length of electrons larger than the widely
accepted coherence length.

Here I propose an alternative explanation of the
magneto-oscillations \cite{ley,ban,sin,proust} as emerging from the
quantum interference of the vortex lattice and the checkerboard or
lattice modulations of the order parameter observed by STM with
atomic resolution \cite{stm}. The checkerboard effectively pins the
vortex lattice, when the period of the latter, $\lambda$ is
commensurate with the period of the checkerboard
 lattice, $a$. The condition $\lambda= Na$ , where $N$ is
a large integer, yields $1/B^{1/2}$ periodicity of the response
functions, rather than $1/B$ periodicity of conventional normal
state
 magneto-oscillations.
\begin{figure}
\begin{center}
\includegraphics[angle=-90
,width=0.90\textwidth]{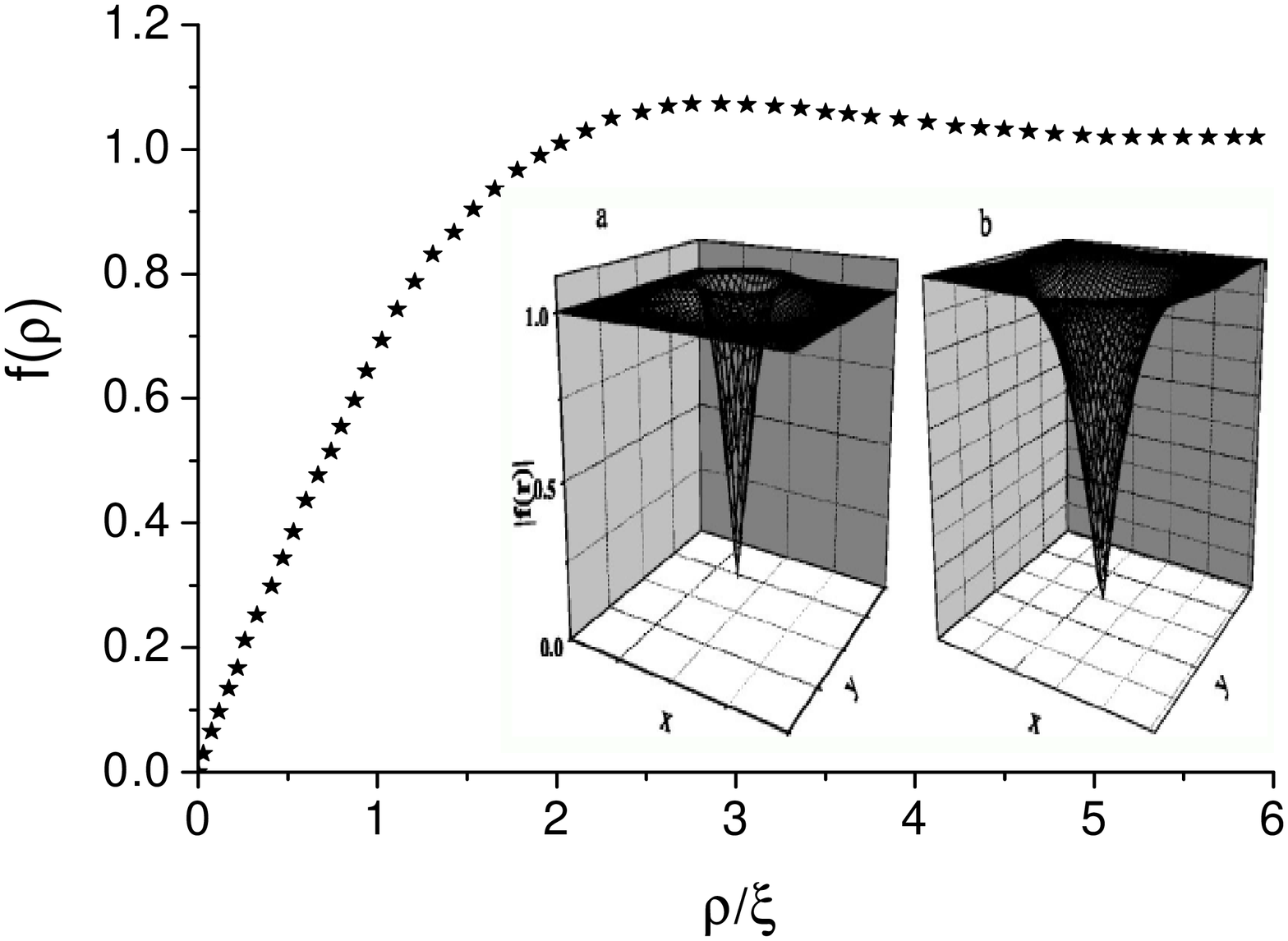} \vskip -0.5mm \caption{The order
parameter profile $f(\rho)=\psi({\bf r})/ n_s^{1/2}$ of a single
vortex in CBL \cite{alevor} (symbols). Inset: CBL vortex (a)
\cite{alevor,alekab} compared with the Abrikosov vortex (b)
\cite{abr} (here $\rho=[x^2+y^2]^{1/2}$). }
\end{center}
\end{figure}

 The integro-differential
equation (\ref{gp})
 in  the
continuum  approximation, $E({\bf K})= \hbar^2K^2/2m$, with the
long-range Coulomb repulsion between double charged bosons, $V({\bf
r})=V_c({\bf r})=4e^2/\epsilon_0 r$,  describes a single vortex with
a \emph{charged} core, Fig.3, when the magnetic field, $B$ is
applied. The coherence
 length in this case is roughly the same as the screening radius, $
 \xi=(\hbar/2^{1/2}m\omega_{p})^{1/2}$.
  Here
$\omega_{p}=(16\pi n_s e^{2}/\epsilon_0 m)^{1/2}$ is the CBL plasma
frequency,  $\epsilon_0$  the static dielectric constant of the host
lattice, $m$ is  the (in-plane) boson mass, and $n_s$ is the average
condensate density.  The chemical potential is zero, $\mu=0$, if one
takes into account the Coulomb interaction alone due to a
neutralizing homogeneous charge background. Each vortex carries one
flux quantum, $\phi_0=\pi \hbar/e$ , but it has an unusual core,
Fig.3a, Ref. \cite{alevor}, due to a local charge redistribution
caused by the magnetic field, different from the conventional vortex
\cite{abr}, Fig.3b. Remarkably,  the coherence length turns out very
small, $\xi \approx 0.5$nm with the material parameters typical for
underdoped cuprates, $m=10 m_{e}$, $n_s=10^{21}cm^{-3}$ and
$\epsilon_{0} = 100$.

The coherence length $\xi$ is so small at low temperatures, that the
distance between two vortices remains large compared with the vortex
size, $\lambda \gg \xi$,  in any  laboratory field reached so far
\cite{ley,ban,sin,proust}. It allows us to write down the
vortex-lattice order parameter, $\psi({\bf r})=\psi_{vl}({\bf r})$,
as
\begin{equation}
\psi_{vl}({\bf r})\approx n_s^{1/2} \left[1- \sum _{j} \phi({\bf r -
r}_j)\right],
\end{equation}
where $\phi({\bf r})=1-f(\rho)$, and  ${\bf r}_j=\lambda\{n_x,n_y\}$
with $n_{x,y}=0, \pm 1, \pm2, ...$ (if, for simplicity,  we take the
square vortex lattice \cite{ref}). The function $\phi(\rho)$ is
linear well inside the core, $\phi(\rho)\approx 1- 1.52 \rho/\xi$
$(\rho\ll \xi)$,  and it has a small negative tail,
$\phi(\rho)\approx -4\xi^4/\rho^4$  outside the core when $\rho
\gg\xi$, Fig.3 \cite{alevor}.

In the continuum approximation with the Coulomb interaction alone
the magnetization of CBL follows the standard logarithmic law, $M(B)
\propto \ln 1/B$ without any oscillations since  the magnetic field
profile is the same as in the conventional vortex lattice
\cite{alekab}. However, more often than not the center-of-mass Bloch
band of preformed pairs, $E({\bf K}),$ has its minima at some finite
wave vectors ${\bf K}={\bf G}$ of their center-of-mass Brillouin
zone \cite{ale5,alebook}. Near the minima the GP equation (\ref{gp})
is written as
\begin{eqnarray}
\left[{(-i\hbar {\bf \nabla}-\hbar{\bf G}+2e {\bf
A})^2\over{2m^{**}}}-\mu \right]\psi({\bf r})+ \cr \int d{\bf r'}
V({\bf r} -{\bf r'})|\psi ({\bf r'})|^2 \psi({\bf r}) =0
\label{gp2},
\end{eqnarray}
with the solution $\psi({\bf r})=\psi_{\bf G}({\bf r})\equiv e^ {i
{\bf G \cdot r}}\psi_{vl}({\bf r})$, if the interaction is the
long-range Coulomb one, $V({\bf r})=V_c({\bf r})$.
\begin{figure}
\begin{center}
\includegraphics[angle=-90
,width=0.90\textwidth]{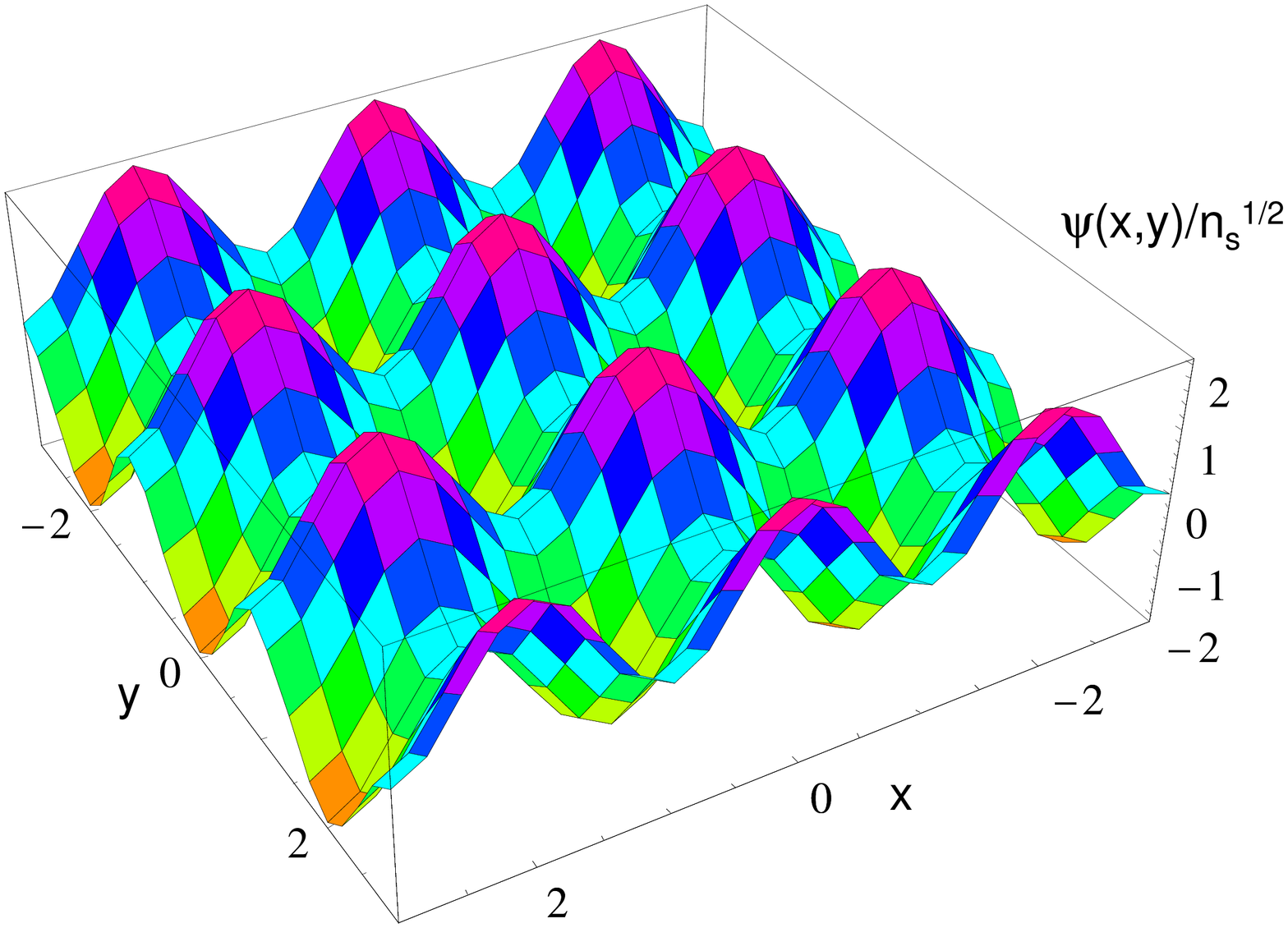} \vskip -0.5mm \caption{The
checkerboard d-wave order parameter of  CBL \cite{alesym} on the
square lattice in zero magnetic field (coordinates $x,y$ are
measured in units of $a$). }
\end{center}
\end{figure}

In particular, a nearest-neighbor (nn)  approximation for the
hopping of intersite bipolarons between oxygen p-orbitals on the
CuO$_2$ 2D lattice yields four generate states $\psi_{\bf G}$ with
${\bf G}_{i} =\{\pm 2\pi/a_0, \pm 2\pi/a_0\}$ , where $a_0$ is the
lattice period \cite{alekor}. Their positions in the Brillouin zone
move towards $\Gamma$ point beyond the nn approximation. The true
ground state is a superposition of four degenerate states,
respecting time-reversal and parity symmetries \cite{alesym},
\begin{equation}
\psi({\bf r})=An_{s}^{1/2}\left[ \cos (\pi x/a)\pm \cos (\pi
y/a)\right]\psi_{vl}({\bf r}). \label{two}
\end{equation}
Here we use the reference frame with $x$ and $y$ axes along the
nodal directions and  $a=2^{-3/2} a_0$. Two "plus/minus" coherent
states, Eq.(\ref{two}), are physically identical since they are
related via a translation transformation, $y \Rightarrow y+ a$.
Normalizing the order parameter by its average value $\langle
\psi({\bf r})^2\rangle=n_s$ and using $(\xi/\lambda)^2 \ll 1$ as a
small parameter  yield the following "minus" state amplitude,
$A\thickapprox 1-N\sum^{\infty}_{n=0}2
[\tilde{\phi}_1(2^{1/2}\pi/a)+\tilde {\phi}_2(2^{1/2}\pi/a)]
\delta_{n, R/2}+ [ \tilde {\phi}_1(2\pi/a)+\tilde {\phi}_2(2\pi/a)]
\delta_{n, R}$ for the square vortex lattice \cite{ref} with the
reciprocal vectors ${\bf g}=(2\pi/\lambda) \{n_x,n_y\}$. Here
$\delta_{n,R}$ is the Kroneker symbol, $R=\lambda/a$ is the ratio of
the vortex lattice period to the checkerboard period
($n=0,1,2,...$),  $N=BS/\phi_0$ is the number of flux quanta in the
area $S$ of the sample, and $\tilde{\phi}_k (q)= (2\pi/S)
\int_0^{\infty} d{\bf \rho} \rho J_0(\rho q) \phi^{k} (\rho) $ is
the Fourier transform of $k$'s power of $\phi(\rho)$, where $J_0(x)$
is the zero-order Bessel function.

The order parameter $\psi({\bf r})$, Eq.(\ref{two}) has the $d$-wave
symmetry changing sign in  real space, when the lattice is rotated
by $\pi /2$. This symmetry is
due to the pair center-of-mass  energy dispersion with the four minima at ${\bf %
K} \neq 0$, rather than due to  a  specific symmetry of the pairing
potential. It also reveals itself as a {\it checkerboard} modulation
of the carrier density with two-dimensional patterns  in zero
magnetic field,  Fig.4, as predicted by us \cite{alesym} prior to
their observations  \cite{stm}. Solving the Bogoliubov-de Gennes
equations with the  order parameter, Eq.(\ref{two}),  yields
 the real-space checkerboard
modulations of the single-particle density of states \cite{alesym},
similar to those observed by STM in cuprate superconductors.

Now we take into account that the interaction between composed pairs
includes a short-range  repulsion along with the long-range Coulomb
one, $V({\bf r})=V_c({\bf r})+v\delta({\bf r})$ \cite{alebook}. At
sufficiently low  carrier density the short-range repulsion can be
treated as a  perturbation to the ground state, Eq.(\ref{two}).
Importantly the short-range repulsion energy of CBL, $U=(v/2)
\langle \psi({\bf r})^4 \rangle$, has a part, $\Delta U $,
oscillating with the magnetic field  as
\begin{equation}
{\Delta U\over{U_0}}\approx N  \sum_{n=0}^{\infty} \left[A_1
\delta_{n,R/2}+ A_2  \delta_{n,R}+A_3
\delta_{n,2R}\right],\label{part}
\end{equation}
where $U_0=v n_s^2/2$ is the hard-core energy of a homogeneous CBL,
and  the amplitudes are proportional to the Fourier transforms of
$\phi(\rho)$ as
\begin{eqnarray}
&&A_1=15 \tilde{\phi}_1\left(2^{1/2}\pi/a\right)- 45
\tilde{\phi}_2\left(2^{1/2}\pi/a\right) +24
\tilde{\phi}_3\left(2^{1/2}\pi/a\right) -
6\tilde{\phi}_4\left(2^{1/2}\pi/a\right)\cr \nonumber
&+&
8\tilde{\phi}_1\left(10^{1/2}\pi/a\right)-12\tilde{\phi}_2\left(10^{1/2}\pi/a\right)+
 8\tilde{\phi}_3\left(10^{1/2}\pi/a\right)
-2\tilde{\phi}_4\left(10^{1/2}\pi/a\right),\\
\end{eqnarray}
 \begin{eqnarray}
&& A_2= -(23/2) \tilde{\phi}_1\left(2\pi/a\right)+
 (57/2)
\tilde{\phi}_2\left(2\pi/a\right) -
16\tilde{\phi}_3\left(2\pi/a\right) +
 4 \tilde{\phi}_4\left(2\pi/a\right)\cr \nonumber
&-&12\tilde{\phi}_1\left(2^{3/2}\pi/a\right)+9\tilde{\phi}_2\left(2^{3/2}\pi/a\right)-
6\tilde{\phi}_3\left(2^{3/2}\pi/a\right)+3\tilde{\phi}_4\left(2^{3/2}\pi/a\right)\\
 \end{eqnarray}
 \begin{equation}
 A_3= -
\tilde{\phi}_1\left(4\pi/a\right)+(3/2)
\tilde{\phi}_2\left(4\pi/a\right) -
\tilde{\phi}_3\left(4\pi/a\right)+
(1/4)\tilde{\phi}_4\left(4\pi/a\right).
\end{equation}

Fluctuations of the pulsed magnetic field and  unavoidable disorder
in cuprates  induce some random distribution of the vortex-lattice
period, $\lambda$. Hence one has to average $\Delta U$ over $R$ with
the Gaussian distribution, $G(R)= \exp
[-(R-\bar{R})^2/\gamma^2]/\gamma \pi^{1/2}$ around an average
$\bar{R}$ with the width $\gamma \ll \bar{R}$. Then  using the
 Poisson summation formula  yields
\begin{eqnarray}
{\Delta U \over{U_0}}= N \sum_{k=0}^{\infty} A_1
e^{-\pi^2k^2\gamma^2/16} \cos (\pi k\bar{R})\cr + A_2
e^{-\pi^2k^2\gamma^2/4} \cos (2\pi k\bar{R})+A_3
e^{-\pi^2k^2\gamma^2} \cos (4\pi k\bar{R}).\label{part2}
\end{eqnarray}

The oscillating correction to the magnetic susceptibility, $\Delta
\chi (B)=-\partial ^2 \tilde{\Omega}/\partial B^2$, is strongly
enhanced due to high oscillating frequencies in Eq.(\ref{part2}).
Since the superfluid has no entropy we can use $\Delta U$ as the
quantum correction to the thermodynamic potential $\tilde{\Omega}$
even at finite temperatures below $T_c(B)$. Differentiating twice
the first harmonic ($k=1$) of the first lesser damped term in
Eq.(\ref{part2}) we obtain
\begin{equation}
\Delta \chi(B) \approx \chi_0  e^{-\delta^2
B_0/16B}\left({B_0\over{B}}\right)^2 \cos (B_0/B)^{1/2},\label{osc}
\end{equation}
where $\chi_0= U_0 SA_1 e^2a^2/4\pi^4\hbar^2$ is a
temperature-dependent amplitude, proportional to the condensate
density squared, $B_0=\pi^3 \hbar/ea^2= 8\pi^3 \hbar/ea_0^2$ is a
characteristic magnetic field, which is approximately $1.1 \cdot
10^6$ Tesla for $a_0 \approx 0.38$ nm, and $\gamma$ is replaced by
$\gamma \equiv \delta \bar{R}$ with the relative distribution width
$\delta$. Assuming that $\xi \gtrsim a $, so that  the amplitude
$A_1$  is roughly $a^2/S$, the quantum correction $\Delta \chi$,
Eq.(\ref{osc}), is of the order of $wx^2/B^2$, where $x$ is the
density of holes per unit cell. It is smaller than the conventional
normal state (de Haas-van Alphen) correction, $\Delta \chi_{dHvA}
\sim \mu/B^2$ \cite{schoen},  for a comparable Fermi-energy scale
$\mu=wx$, since $x \ll 1$ in the underdoped cuprates.

\begin{figure}
\begin{center}

\includegraphics[angle=-90
,width=0.80\textwidth]{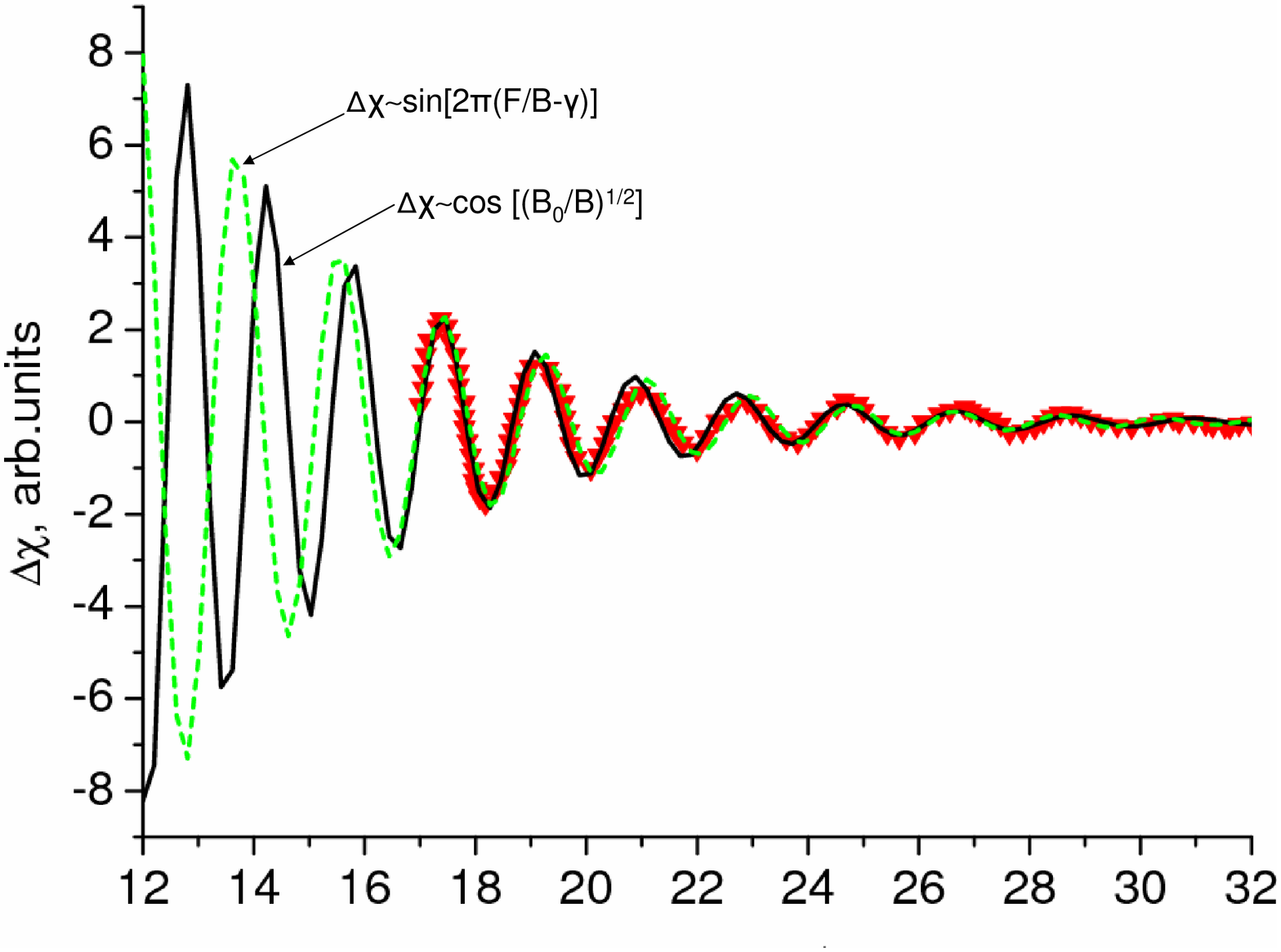}
 \caption{Quantum corrections to
 the vortex-lattice  susceptibility versus $1/B$, Eq.(\ref{osc}) (solid line,
$B_0=1.000\cdot 10^6$ Tesla, $\delta=0.06$) compared with
oscillating susceptibility of YBa$_2$Cu$_3$O$_{6.5}$ (symbols) and
with the conventional normal state oscillations (dashed line)
\cite{proust} at $T=0.4$ K.}
\end{center}
\end{figure}
Different from normal state dHvA oscillations, which are periodic
versus $1/B$, the vortex-lattice oscillations, Eq.(\ref{osc}) are
periodic versus $1/B^{1/2}$. They are quasi-periodic versus $1/B$
with a field-dependent  frequency $F=B_0 (B/B_0)^{1/2}/2\pi$, which
is strongly reduced relative to the conventional-metal frequency
($\approx B_0/2\pi$) since $B\ll B_0$, as observed in the
experiments \cite{ley,ban,sin,proust}. The quantum correction to the
susceptibility, Eq.(\ref{osc})  fits well the oscillations in
YBa$_2$Cu$_3$O$_{6.5}$ \cite{proust}, Fig.5. The oscillations
amplitudes, proportional to $n_s^2\exp(-\delta^2 B_0/16B)$
  decay  with increasing temperature since the randomness of the vortex lattice, $\delta$,
  increases, and
 the Bose-condensate evaporates.

\section{Summary}
A possibility of real-space pairing, proposed originally by Ogg
\cite{ogg} and later on by Schafroth  and Blatt and Butler
\cite{but}   has been the subject of many discussions as opposed to
the Cooper pairing, particularly heated over the last 20 years after
the discovery of high temperature superconductivity in cuprates. Our
extension of the BCS theory towards the strong interaction between
electrons and ion vibrations  \cite{alebook} proved that BCS and
Ogg-Schafroth pictures are two extreme limits of the same problem.
 Cuprates are characterized by poor
screening of high-frequency optical phonons, which allows the
long-range Fr\"ohlich EPI bound holes in  superlight small
bipolarons \cite{ale5}, which are several orders of magnitude
lighter than small Holstein (bi)polarons.

The bipolaron theory accounts for GPE and NPE
 in slightly doped semiconducting and undoped insulating
cuprates, respectively. It predicts  the occurrence of a new length
scale, $\hbar/\sqrt{2m_c\epsilon (T)}$, which turns out much larger
than the zero-temperature coherence length in a wide temperature
range  above the transition point of the normal barrier, if bosons
are almost two-dimensional (2D). The physical reason, why the
quasi-2D bosons display a large normal-state coherence length,
whereas 3D Bose-systems (or any-D Fermi-systems) at the same values
of parameters do not, originates in the large DOS near the band edge
of two-dimensional bosons compared with 3D DOS. Since  DOS is large,
the chemical potential is pinned near the edge with the magnitude,
$\epsilon (T)$, which is  exponentially small

 Here I also propose that the magneto-oscillations in underdoped
cuprate superconductors  result from  the quantum interference of
the vortex lattice and the lattice modulations of the d-wave order
parameter, Fig.4, which play the role of a periodic pinning grid.
Our expression (\ref{osc}) describes the oscillations as well as the
standard Lifshitz-Kosevich formula of dHvA and SdH effects \cite{
ley,ban,sin,proust}. The difference of these two dependencies could
be resolved in ultrahigh magnetic fields as shown in Fig.5. While
our theory  utilizes GP-type equation for composed charged bosons
\cite{alevor}, the
  quantum interference of vortex and crystal
lattice modulations of the order parameter is quite universal
extending well beyond Eq.(\ref{gp}) independent of a particular
pairing mechanism. It can also take place in the standard BCS
superconductivity at $B < H_{c2}$, but  hardly be observed because
of much lower value of $H_{c2}$ in conventional superconductors
resulting in a very small damping factor, $\propto \exp(-\delta^2
B_0/16B) \lll 1$.

I  appreciate valuable discussions with A. F. Andreev, I. Bozovic,
A. Bussmann-Holder, J. T. Devreese, L. P. Gor'kov,  V. V. Kabanov,
and support of this work by EPSRC (UK) (grant Nos. EP/D035589,
EP/C518365).


\begin{chapthebibliography}{1}
\bibitem{alexandrov:1988} A. S. Alexandrov, Phys. Rev. B {\bf 38},
925 (1988).
\bibitem{alemot} A. S.  Alexandrov and N. F. Mott, Rep. Prog. Phys. ${\bf 57}$,
1197 (1994).
\bibitem{dev} J.T. Devreese, in \emph{Encyclopedia of Applied Physics}, vol. 14, p. 383 (VCH Publishers (1996)).
\bibitem{ale5}  A. S. Alexandrov, Phys. Rev. B {\bf 53}, 2863
(1996).
\bibitem{alebook} A.S. Alexandrov,
\emph{Theory of Superconductivity: From Weak to Strong Coupling}
(IoP Publishing, Bristol and Philadelphia, 2003).

\bibitem{edw} P. P. Edwards, C. N. R. Rao, N. Kumar, and A. S.
Alexandrov, ChemPhysChem {\bf 7}, 2015 (2006)

\bibitem{alenar} A. S. Alexandrov, in \emph{Studies in High Temperature
Superconductors,} ed. A.V.  Narlikar (Nova Science Pub., NY 2006),
v.\textbf{50}, pp. 1-69.

\bibitem{alejpcm} A. S. Alexandrov,  J.
Phys.: Condens. Matter {\bf 19}, 125216 (2007).

\bibitem{verbist} G. Verbist,
F.~M. Peeters, and J.~T. Devreese, Phys. Rev. B {\bf 43}, 2712
(1991).

\bibitem{alekor}  A. S. Alexandrov and P. E. Kornilovitch, J. Phys.
Cond. Matt. {\bf 14}, 5337(2002)
\bibitem{hague} J. P. Hague, P. E. Kornilovitch, J. H. Samson, and A. S. Alexandrov
Phys. Rev. Lett. {\bf 98}, 037002 (2007).


\bibitem{zhao}  G. M. Zhao and D. E. Morris Phys. Rev. B {\bf
51}, 16487 (1995); G.-M. Zhao, M. B. Hunt, H. Keller, and K. A.
M\"uller, Nature (London) {\bf
 385}, 236 (1997); R. Khasanov, D. G. Eshchenko, H. Luetkens, E. Morenzoni, T. Prokscha, A. Suter,
 N. Garifianov, M. Mali, J. Roos, K. Conder, and H. Keller,
Phys. Rev. Lett. {\bf 92}, 057602 (2004), A. Bussmann-Holder, H.
Keller, A. R. Bishop, A. Simon, R. Micnas, and K. A. Muller,
EuroPhys. Lett. {\bf 72},  423 (2005).
\bibitem{ref2} Recently D. R. Harshman, J. D. Dow, and A. T. Fiory  (Phys. Rev. B {\bf
77}, 024523 (2008)) have reinterpreted the isotope effects in
cuprates in terms of a pair-breaking mechanism, rather than
polarons.  These authors claim that the superconducting condensate
resides in the buffer layers, but not in the $CuO_2$ planes, and
therefore it has  s-wave, rather than d-wave symmetry. However they
neglect the textbook ("Anderson") theorem by applying the
 magnetic-impurity pair-breaking theory by Abrikosov and
Gor'kov to \emph{nonmagnetic } impurities. Moreover disregarding the
Fr\"ohlich EPI they have also violated the Coulomb law.
\bibitem{LAN}  A. Lanzara, P.V. Bogdanov, X.J. Zhou, S.A.
Kellar, D.L. Feng, E.D. Lu, T. Yoshida, H. Eisaki, A. Fujimori,K.
Kishio, J.I. Shimoyana, T. Noda, S. Uchida, Z. Hussain and Z.X.
Shen, Nature (London) {\bf 412}, 510 (2001); G-H. Gweon, T.
Sasagawa, S.Y. Zhou, J. Craf, H. Takagi, D.-H. Lee, and A. Lanzara,
Nature (London) {\bf 430}, 187 (2004); X. J. Zhou, J. Shi, T.
Yoshida, T. Cuk, W. L. Yang, V. Brouet, J. Nakamura, N. Mannella, S.
Komiya, Y. Ando, F. Zhou, W. X. Ti, J. W. Xiong, Z. X. Zhao, T.
Sasagawa, T. Kakeshita, H. Eisaki, S. Uchida, A. Fujimori, Z.-Y.
Zhang, E. W. Plummer, R. B. Laughlin, Z. Hussain, and Z.-X. Shen,
Phys. Rev. Lett. {\bf 95}, 117001 (2005).
\bibitem{mic1}  D. Mihailovic, C.M. Foster, K. Voss, and A.J. Heeger, Phys. Rev. B {\bf 42}, 7989 (1990).
\bibitem{ita}  P. Calvani, M. Capizzi, S. Lupi, P. Maselli, A. Paolone, P. Roy, S.W. Cheong, W. Sadowski,
and E. Walker, Solid State Commun. {\bf 91}, 113 (1994).
\bibitem{tal} R. Zamboni, G. Ruani, A.J. Pal, and C. Taliani,
Solid St. Commun. {\bf 70}, 813 (1989).
\bibitem{tim} T. Timusk, C.C. Homes,  and W. Reichardt, in
\emph{Anharmonic properties of High Tc cuprates} (eds. D.
Mihailovic, G. Ruani, E. Kaldis, and K.A. M\"uller, Singapore: World
Scientific, p.171 (1995)).
\bibitem{devopt} J. Tempere and J. T. Devreese, Phys. Rev. B {\bf 64}, 104504 (2001).
\bibitem{ega} T.R. Sendyka, W. Dmowski,  T. Egami, N. Seiji, H. Yamauchi, and S. Tanaka,
 Phys. Rev. B {\bf 51}, 6747 (1995);
T. Egami, J. Low Temp. Phys. {\bf 105}, 791 (1996).
\bibitem{rez} D. Reznik, L. Pintschovius, M. Ito, S. Iikubo,
M. Sato, H. Goka, M. Fujita, K. Yamada, G. D. Gu, and J. M.
Tranquada,  Nature {\bf 440}, 1170 (2006).
\bibitem{zha} P. Zhang, S. G. Louie and M. L. Cohen, Phys. Rev. Lett. {\bf 98},
067005 (2007).
\bibitem{alexandrov:2001b} A. S. Alexandrov, Physica C (Amsterdam) {\bf 363}, 231 (2001).

\bibitem{kabmic}  D. Mihailovic, V.V. Kabanov, K. Zagar, and J.
Demsar, Phys. Rev. B{\bf 60}, 6995 (1999) and references therein.
\bibitem{alexandrov:1991} A. S. Alexandrov, Physica C (Amsterdam) {\bf 182}, 327 (1991).
\bibitem{ZAV}  V.N. Zavaritsky, V.V. Kabanov and A.S. Alexandrov, Europhys.
Lett. {\bf 60}, 127 (2002).
\bibitem{aleH} A. S. Alexandrov, Phys. Rev. B {\bf 48},
10571 (1993).
\bibitem{NEV}  A. S. Alexandrov and N. F. Mott,\emph{ Phys. Rev. Lett.} {\bf 71},
1075 (1993).
\bibitem{ZHANG}  Y. Zhang, N. P. Ong,
Z. A. Xu, K. Krishana, R. Gagnon and L. Taillefer, Phys. Rev. Lett.
{\bf 84}, 2219 (2000), and unpublished.
\bibitem{leeale} K. K. Lee, A. S. Alexandrov,  and W. Y. Liang,
 Phys. Rev. Lett. {\bf 90},  217001 (2003); Eur. Phys. J. B {\bf
30}, 459 (2004).
\bibitem{alelor} A. S. Alexandrov, Phys.
Rev. B \textbf{73}, 100501 (2006).
\bibitem{aledia} A. S. Alexandrov,  Phys. Rev. Lett. \textbf{96}, 147003 (2006).
\bibitem{polarons} H. Fehske and S. A. Trugman, in \emph{Polarons in Advanced
Materials}, ed. A. S. Alexandrov (Springer/Canopus, Bristol 2007),
pp.393-461; A. S. Mishchenko and N. Nagaosa, ibid, pp. 503-544.
\bibitem{proxi} A. S. Alexandrov, Phys. Rev. B {\bf 75}, 132501 (2007).
\bibitem{bozp} I. Bozovic, G. Logvenov, M. A. J. Verhoeven, P. Caputo, E. Goldobin, and M. R.
Beasley, Phys. Rev. Lett. \textbf{93}, 157002 (2004), and references
therein.
\bibitem{boz5} P. Abbamonte, L. Venema, A. Rusydi, G. A. Sawatsky, G. Logvenov, and I. Bozovic, Science {\bf 297}, 581 (2002).
\bibitem{boz0}
I. Bozovic, G. Logvenov, M. A. J. Verhoeven, P. Caputo, E. Goldobin,
and  T. H. Geballe,  Nature (London) {\bf 422}, 873 (2003).
\bibitem{alevor}
A. S. Alexandrov, Phys. Rev. B {\bf 60}, 14573 (1999).

\bibitem{gin}
V. L. Ginzburg and L. D. Landau, Zh. Eksp. Teor. Fiz. {\bf 20}, 1064
(1950).
\bibitem{gro}
E. P. Gross, Nuovo Cimento ${\bf 20}$, 454 (1961); L.P. Pitaevskii,
Zh. Eksp. Teor. Fiz. {\bf 40}, 646 (1961) ( Soviet Phys. JETP ${\bf
13}$, 451 (1961)).
\bibitem{kagan} M. Yu. Kagan and D. V. Efremov, Phys. Rev. B {\bf
65}, 195103 (2002).
\bibitem{popov} V. N. Popov, Theor. Math. Phys.
{\bf 11}, 565 (1972); D. S. Fisher and P. C. Hohenberg, Phys. Rev. B
{\bf 37}, 4936 (1988).

\bibitem{mac} A. P. Mackenzie \emph{et
al.}, Phys. Rev. Lett. {\bf 71}, 1238 (1993).
\bibitem{plate} M. Plate \emph{et
al.}, Phys. Rev. Lett. 95, 077001 (2005).
\bibitem{ley} N. Doiron-Leyraud \emph{et
al.}, Nature {\bf 447}, 565 (2007).
\bibitem{ban}  A. F. Bangura \emph{et
al.}, arXiv:0707.4461.
\bibitem{sin}  E. A. Yelland \emph{et
al.}, arXiv:0707.0057.
 \bibitem{proust} C. Jaudet \emph{et
al.}, arXiv:0711.3559.
\bibitem{shen} A. Damascelli, Z. Hussain and
Zhi-Xun Shen, Rev. Mod. Phys. {\bf 75} 473 (2003).
\bibitem{schoen}
D. Schoenberg, \emph{Magnetic Oscillations in Metals} (Cambridge
University Press, Cambridge 1984).

\bibitem{harris} J. M. Harris, K. Krishana, N. P. Ong, R Cagnon, and
L. Taillefer, J. Low Temp. Phys. {\bf 105}, 877 (1996).

\bibitem{stm}  J. E. Hoffman \emph{et
al.}
  Science {\bf 295}, 466 (2002);
C. Howald \emph{et al.}, Phys. Rev. B{\bf 67}, 014533 (2003); M.
Vershinin \emph{et al.} Science {\bf 303}, 1995 (2004).

\bibitem{abr}
A. A. Abrikosov, Zh. Eksp. Teor. Fiz. {\bf 32}, 1442 (1957) ( Soviet
Phys. JETP ${\bf 5}$, 1174 (1957)).
\bibitem{alekab} V. V. Kabanov and A. S. Alexandrov, Phys. Rev. B
{\bf 71}, 132511 (2005).
\bibitem{ref} Results for the square vortex lattice are also applied to the
triangular lattice. Moreover there is  a crossover from triangular
to square coordination of vortices with increasing magnetic field in
the mixed phase of cuprate superconductors (R. Gilardi \emph{et
al.}, Phys. Rev. Lett. {\bf 88}, 217003 (2002); S. P. Brown \emph{et
al.}, Phys. Rev. Lett. 92, 067004 (2004) ).
\bibitem{alesym} A. S.
Alexandrov, Physica C {\bf 305}, 46 (1998); Int. J.  Mod. Phys. B
{\bf  21},  2301 (2007).

\bibitem{ogg} R. A. Ogg Jr., Phys. Rev. {\bf 69}, 243
(1946).
\bibitem{but}M. R. Schafroth, Phys. Rev. {\bf 100}, 463 (1955);  J. M.  Blatt and S.T. Butler,  Phys. Rev. {\bf %
100}, 476 (1955).

\end{chapthebibliography}
\end{document}